\documentclass[aps,prb,amsmath,amssymb,reprint,superscriptaddress]{revtex4-2}

\usepackage[colorlinks,citecolor=blue,linkcolor=red,urlcolor=blue]{hyperref}
\usepackage{color}
\usepackage{graphicx}
\usepackage{subfigure}
\usepackage{verbatim}
\usepackage{fourier}
\usepackage{array}
\usepackage{hhline}
\usepackage{multirow}

\begin{document}

\preprint{HIT-L2C-UM-CNRS}

\title{Photothermal behavior for two-dimensional nanoparticle ensembles: \\multiple scattering and thermal accumulation effects}

\author{Minggang Luo}
%\email[]{luohiter@gmail.com(Minggang Luo)}
\affiliation{School of Energy Science and Engineering, Harbin Institute of Technology, 92 West Street, Harbin 150001, China}
\affiliation{Laboratoire Charles Coulomb (L2C) UMR 5221 CNRS-Universit\'e de Montpellier, F- 34095 Montpellier, France}

\author{Junming Zhao}
\email[]{jmzhao@hit.edu.cn(Junming Zhao)}
\affiliation{School of Energy Science and Engineering, Harbin Institute of Technology, 92 West Street, Harbin 150001, China}
\affiliation{Key Laboratory of Aerospace Thermophysics, Ministry of Industry and Information Technology, Harbin 150001, China}

\author{Linhua Liu}
\affiliation{School of Energy and Power Engineering, Shandong University, Qingdao 266237, China}

\author{Mauro Antezza}
\email[]{mauro.antezza@umontpellier.fr(Mauro Antezza)}
\affiliation{Laboratoire Charles Coulomb (L2C) UMR 5221 CNRS-Universit\'e de Montpellier, F- 34095 Montpellier, France}
\affiliation{Institut Universitaire de France, 1 rue Descartes, F-75231 Paris Cedex 05, France}

\date{\today}

\begin{abstract}

Light-assisted micro-nanoscale temperature control in complex nanoparticle network attracts lots of research interests. Many efforts have been put on the optical properties of the nanoparticle networks and only a few investigations on its light-induced thermal behavior was reported. We consider two-dimensional (2D) square-lattice nanoparticle ensemble made of typical metal Ag with a radius of 5 nm. The effect of complex multiple scattering and thermal accumulation on the light-induced thermal behavior in plasmonic resonance frequency (around 383 nm) is analyzed through the Green\textquotesingle s function approach. Regime borders of both multiple scattering and thermal accumulation effects on the photothermal behavior of 2D square-lattice nanoparticle ensemble are figured out clearly and quantitatively. A dimensionless parameter $\varphi$ is defined as the ratio of full temperature increase to that without considering the multiple scattering or thermal accumulation to quantify the multiple scattering and thermal accumulation effects on photothermal behavior. The more compact the nanoparticle ensemble is, the stronger the multiple scattering on thermal behavior is. When the lattice spacing increases to tens of nanoparticle radius, the multiple scattering becomes insignificant. When $\varphi \approx 1$ (lattice spacing increases to hundreds of nanoparticle radius), the thermal accumulation effects are weak and can be neglected safely. The polarization-dependent distribution of temperature increase of nanoparticles is observed only in the compact nanoparticle ensemble, while for dilute ensemble, such polarization-dependent temperature increase distribution can not be observed anymore. This work may help for the understanding of the light-induced thermal transport in the 2D particle ensemble.

\end{abstract}

\maketitle 

\section{Introduction}
Nanoscale temperature control in complex plasmonic nanoparticle ensembles attracts a lot of interest in the fields of physics \cite{Baffou2020NM,Baffou2020LSA,Blum2015,Maier2001}, chemistry \cite{Li2019,Zhou2018,Saha2012,Kamat2002}, and biology \cite{Michael2018,Hana2018,Jones2018}, to name a few, where the light illumination is an efficient and common ingredient. Such a light-assisted temperature control approach has many applications, ranging from hyperthermia therapy to additive manufacturing. Nanoparticles inside the ensemble often obtain energy from the incident light and then work as heat sources heating each other. Hence, the photothermal behavior for nanoparticle ensemble is a coupling optical-thermal process and should be analysed from both optical (light absorption and scattering) and thermal (heat dissipation) sides.

To well understand the light-induced thermal behavior of nanoparticle ensembles, we should investigate the optical properties of the nanoparticle ensembles at first. Multiple scattering (MS) in nanoparticle ensembles will inevitably affect light absorption and finally affect the light-induced thermal behavior, which is named the multiple scattering effect. The coupled dipole method (CDM) is a well-known tool used for the investigation on the properties of small particles \cite{Merchiers2007PRA,Mulholland1994}. In CDM, the effect of mutual multiple scattering in the nanoparticle ensemble on the light-absorption is considered by using the external electric field experienced by each nanoparticle rather than the direct incident field (i.e., illumination field). When nanoparticles are far enough away from each other, hence, they can be treated as optically independent. That is the external electric field experienced by nanoparticles can directly be approximated to the incident field. Some other methods (e.g., finite element method \cite{Borah2019} and the finite difference time domain method \cite{Alexander2020}, generalized multiparticle Mie method \cite{Xu2003,Khlebtsov2006} and boundary element method \cite{Abajo1998PRL,Abajo2002PRB,Baffou2010ACSnano}, to name a few) can also be applied to investigate the optical properties of nanoparticle ensembles.

Then, with the aforementioned methods in hand, the optical properties of nanoparticle ensembles have been investigated widely and have been reviewed in the literature \cite{Abajo2007RMP,Ross2016CRL,Kravets2018CLR}. Collective lattice resonance for disordered and quasi-random ensembles \cite{Vadim2019JOSAB}, and ordered arrays \cite{Utyushev2020,Manjavacas2019ACSnano,Zakomirnyi2019OL,Ramezani2016prb,Rodriguez2012} of nanoparticles were analyzed. The effect of array structure on plasmonic resonance wavelength was analyzed theoretically for the silver nanoparticle ensembles \cite{Zhao2003}, where extinction spectra shift was reported. Evlyukhin \textit{et al}. \cite{Evlyukhin2010prb} and Zundel \textit{et al}. \cite{Zundel2018} systematically analyzed the finite-size effect on the optical response for periodic arrays of nanostructures (e.g., nanoparticles and graphene nanodisks). Besides the above theoretical predictions, electromagnetic interactions in plasmonic nanoparticle ensembles were investigated and the spectra shift was demonstrated experimentally \cite{Bouhelier2005}. The aforementioned investigations on the optical response of nanoparticle ensembles provide the fundamental knowledge from the optical side to understand the photothermal behavior of nanoparticle ensembles.

From the thermal side, light absorption by each nanoparticle in the ensemble will heat the whole nanoparticle ensemble cumulatively, namely accumulative (collective) heating effect, which was predicted theoretically \cite{Baffou2013singleNP,Un2019JAP,Baffou2010,Gillibert2020,Heber2014AcsNano} and demonstrated experimentally \cite{Richardson2009NL,Baffou2013AcsNano,Baffou2014Nanoscale,Constantinos2019JPCC}. For a small particle (point dipole approximation valid), a Green\textquotesingle s function approach combining the coupled dipole method with the thermal Green\textquotesingle s function together for the light-induced thermal behavior modeling of nanoparticle ensembles was proposed by Baffou \textit{et al}. (2010) \cite{Baffou2010}, where the steady-state temperature distribution throughout arbitrary complex plasmonic systems can be calculated easily. 

By means of the Green\textquotesingle s function approach, the effect of plasmonic (optical) coupling on the photothermal behavior of an ensemble with three-dimensional (3D) random distribution of nanoparticles was analyzed by Siahpoush \textit{et al}. \cite{Siahpoush2018OC}. They found that the multiple scattering can reduce the temperature increase of nanoparticles as compared to the case without multiple scattering at the plasmonic resonance wavelength of a single nanoparticle. Due to the thermal accumulation (TA) effect, when calculating the temperature increase of an arbitrary nanoparticle from a nanoparticle ensemble, we often can\textquotesingle t simply treat the nanoparticle solely, just like the rest nanoparticles don\textquotesingle t exist \cite{Govorov2006NRL,Baffou2013AcsNano}. Though the multiple scattering and thermal accumulation on the photothermal behavior of nanoparticle ensembles with different configurations have already been investigated (as summarized in Table \ref{summary}) , no clear and quantitative regime borders of these two effects have been reported yet. The method to figure out the regime of these two effects will greatly facilitate the investigation of the photothermal behavior of nanoparticle ensembles, which is the motivation of this work.

\begin{table}[htbp]
  \caption{Different configurations in the literature in the field of thermal behavior of nanoparticle ensemble under illumination}
%  \centering
  \begin{tabular}{|l|l|l|}
   \hline
  %\toprule
     Year & Authors & Configurations\\
   \cline{1-3}
   \hhline{|===|}
   \multirow{2}*{2006}&\multirow{2}*{Govorov \textit{et al}. \cite{Govorov2006NRL}} & 1NP, 2NPs, 2D 4$\times$4 square-lattice\\
    & & NP ensembles\\
   \cline{1-3}
   2009 & Richardson \textit{et al}. \cite{Richardson2009NL} & Metal NP solution\\
   \cline{1-3}
   2010 & Baffou \textit{et al}. \cite{Baffou2010} & 2D square-lattice NP ensembles\\
   \cline{1-3}
   \multirow{2}*{2013}&\multirow{2}*{Baffou \textit{et al}. \cite{Baffou2013AcsNano}} & 1D linear NP chain, 2D square-lattice\\
    & & NP ensembles\\
   \cline{1-3}
  2018 & Siahpoush \textit{et al}. \cite{Siahpoush2018OC} &  3D random NP ensembles\\
   \cline{1-3}
  2019 & Moularas \textit{et al}. \cite{Constantinos2019JPCC} &  Core-shell NP ensembles\\
   \cline{1-3}
   
%\bottomrule
  \end{tabular}
  \label{summary}
 \end{table}

It should be noted that, due to the multiple scattering, the light may attenuate along its propagation direction in the 3D random nanoparticle ensemble, which can also reduce the temperature increase of nanoparticles. It is hard to tell the effect of light attenuation on temperature increase from the multiple scattering on inhibition of the temperature. In this work, to have a clear understanding of the multiple scattering effect on photothermal behavior without any interference of the light attenuation, we consider the two-dimensional nanoparticle ensembles, of which the extension direction is perpendicular to the light propagation direction. In addition, light absorption of nanoparticle ensemble is polarization-dependent \cite{Ma2020AO,Borah2019,Ross2016CRL}. The effect of light polarization on the photothermal behavior of an ensemble composed of only a few or tens of nanoparticles has been analyzed already \cite{Borah2019,Ren2018IJHMT,Metwally2017,Baffou2010ACSnano}. As the number of nanoparticles in the ensemble increases, the mutual interaction between nanoparticles in the ensemble becomes more and more complex, of which it\textquotesingle s remained unclear if the light polarization effect on the photothermal behavior still exists.

We address the aforementioned missing points in this paper, where the photothermal behavior of two-dimensional nanoparticle ensembles is investigated using the Green\textquotesingle s function approach \cite{Baffou2010}. This work is organized as follows. In Sec.~\ref{Theorectical_aspect}, the coupled dipole method for light scattering and the thermal Green\textquotesingle s function method for steady temperature spatial distribution are presented in brief, with considering the mutual multiple scattering interaction and thermal accumulation effect. In addition, the physical model of the 2D nanoparticle ensemble considered in this work is also given. In Sec.~\ref{Results}, a method to clearly and quantitatively figure out the regime borders of both the multiple scattering and the thermal accumulation will be proposed as the focus of this work. Effects of both the multiple scattering and the thermal accumulation on photothermal behavior, as well as the relation between the two effects, will be discussed.

\section{Theoretical models}
\label{Theorectical_aspect}
In this section, we describe the physical system and the theoretical models: (1) the coupled dipole method describing the light scattering and absorption of nanoparticle ensembles and (2) thermal Green\textquotesingle s function method describing the steady temperature spatial distribution for the nanoparticle ensembles. The SI unit system is used.

\subsection{Physical systems: 2D nanoparticle ensembles}
\begin{figure} [htbp]
\centerline {\includegraphics[scale=0.5]{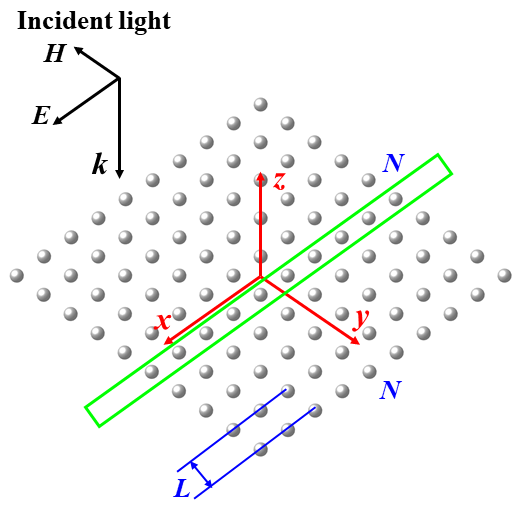}}
        \caption{Structure diagram of the two-dimensional $N\times N$ square-lattice nanoparticle ensemble. Light-induced thermal behavior of the ensemble is investigated. The lattice spacing is $L$. Nanoparticle radius is $a$. A chain of nanoparticles near the $x$-axis is extracted out of the 2D square-lattice nanoparticle ensemble in the green frame and is defined by $y=L/2$ and $z=0$.}
        \label{structure}
\end{figure}

We investigate the thermal behavior of the 2D nanoparticle ensemble ($N\times N$ square-lattice ensemble) illuminated by an incident light, as shown in Fig.~\ref{structure}. The size of the nanoparticle is supposed to be small compared to the wavelength of the incident light, which results in the validity of the point dipole approximation of the nanoparticle. The incident light wave vector is against the positive $z$-axis direction and the polarization direction is parallel to the $x$-axis. The ensemble is parallel to the $xoy$ plane. The lattice spacing is $L$. Nanoparticle radius is $a$. When analyzing the relation between the multiple scattering and collective effects on the light-induced thermal behavior of the ensemble, a chain of nanoparticles of interest near the $x$-axis is extracted out of the 2D square-lattice nanoparticle ensemble in the green frame and is defined by $y=L/2$ and $z=0$. When analyzing the thermal accumulation effect, the nanoparticle in the first sector closest to the origin is assigned as the central nanoparticle of the ensemble. The central nanoparticle position is defined as $x=y=L/2$.

\subsection{Coupled dipole method}
\label{DDA}
When the dipolar nanoparticle ensemble is illuminated by an incident light characterized by the electric field amplitude $\textbf{E}^{\rm inc}(\textbf{r})$, the multiple scattering between each nanoparticle results in an external field  $\textbf{E}^{\rm ext}(\textbf{r})$ experienced by each nanoparticle, which is quite different from the incident one and yields \cite{Mulholland1994,Zundel2018}
\begin{equation}
\textbf{E}^{\rm ext}(\textbf{r}_{\rm i})=\textbf{E}^{\rm inc}(\textbf{r}_{\rm i})+\frac{k^2}{\epsilon_0\epsilon_{\rm m}}\sum_{{\rm j}\neq {\rm i}}^{N_{\rm t}}\textbf{G}(\textbf{r}_{\rm i},\textbf{r}_{\rm j})\cdot\textbf{p}_{\rm j},
\label{E_ext}
\end{equation}
where $k=\sqrt{\epsilon_{\rm m}}\omega/c$ is the wave vector in the host medium, $\epsilon_{\rm m}$ is the host medium relative permittivity, $\epsilon_0$ is vacuum permittivity, $N_{\rm t}=N\times N$ is the total number of nanoparticles in the 2D square-lattice nanoparticle ensemble, $G(\textbf{r}_{\rm i},\textbf{r}_{\rm j})=\frac{e^{ikr}}{4\pi r}\left[\left(1+\frac{ikr-1}{k^{2}r^{2}}\right)\mathbb{I}_{3}+\frac{3-3ikr-k^{2}r^{2}}{k^{2}r^{2}}\hat{\textbf{r}}\otimes\hat{\textbf{r}}\right]$ is the electric Green\textquotesingle s function connecting two nanoparticles at $\textbf{r}_{\rm i}$ and $\textbf{r}_{\rm j}$, $r$ is the magnitude of the separation vector $\textbf{r}=\textbf{r}_{\rm i}-\textbf{r}_{\rm j}$, $\hat{\textbf{r}}$ is the unit vector $\textbf{r}/r$, $\mathbb{I}_{3}$ is the $3\times3$ identity matrix, $\textbf{p}_{\rm j}$ is the dipole moment located at $\textbf{r}_{\rm j}$, which yields
\begin{equation}
\textbf{p}_{\rm j} = \epsilon_0 \epsilon_{\rm m} \alpha_{\rm j} \textbf{E}^{\rm ext}(\textbf{r}_{\rm j}),
\label{dipole moment}
\end{equation}
where $\alpha_{\rm j}$ is the polarizability of the $\rm j$th particle considering the radiation correction defined as follows.
\begin{equation}
\alpha_{\rm j}=\left(\frac{1}{\alpha_{\rm j}^0}-\frac{ik^3}{6\pi}\right)^{-1},
\label{a_E}
\end{equation}
where $\alpha_{\rm j}^0$ is the Clausius-Mossotti polarizability defined as \cite{Chapuis2008,Kravets2018CLR}
\begin{equation}
\alpha_{\rm j}^{0}=4\pi a^3\frac{\epsilon (\omega)-\epsilon_{\rm m}}{\epsilon (\omega)+2\epsilon_{\rm m}},
\label{a_E_CM}
\end{equation}
where $a$ is the nanoparticle radius. $\epsilon(\omega)$ is the relative permittivity of the nanoparticle. Then, the external field experienced by all nanoparticles in the ensemble can be rearranged in a compact and explicit way as follow:
\begin{equation}
\begin{pmatrix}
\textbf{E}^{\rm ext}(\textbf{r}_{\rm 1})\\
\textbf{E}^{\rm ext}(\textbf{r}_{\rm 2})\\
\vdots\\
\textbf{E}^{\rm ext}(\textbf{r}_{\rm N_{\rm t}})
\end{pmatrix}=\mathbb{A}^{-1}\begin{pmatrix}
\textbf{E}^{\rm inc}(\textbf{r}_{\rm 1})\\
\textbf{E}^{\rm inc}(\textbf{r}_{\rm 2})\\
\vdots\\
\textbf{E}^{\rm inc}(\textbf{r}_{\rm N_{\rm t}})
\end{pmatrix},
\label{E_ext_compact}
\end{equation} 
where the $3N\times 3N$ matrix is defined as follows.
\begin{equation}
\mathbb{A}=\mathbb{I}_{3N}^{}-k^2\begin{pmatrix}
0 & \alpha_{2}\textbf{G}_{12} & \cdots & \alpha_{N}\textbf{G}_{1N}\\
\alpha_{1}\textbf{G}_{21} & 0 & \ddots & \vdots\\
\vdots & \vdots & \ddots & \alpha_{N}\textbf{G}_{(N-1)N}\\
\alpha_{1}\textbf{G}_{N1} & \cdots & \alpha_{N-1}\textbf{G}_{N(N-1)} & 0
\end{pmatrix},
\label{matrix_interaction}
\end{equation} 
where $\textbf{G}_{\rm ij}=\textbf{G}(\textbf{r}_{\rm i},\textbf{r}_{\rm j})$, $\mathbb{I}_{3N}$ is the $3N\times 3N$ identity matrix.

\subsection{Thermal Green\textquotesingle s function method}
\label{TDDA}
In the following, it is supposed that the thermal conductivity of the nanoparticle is much higher than the one of the host medium. Under such approximation, the temperature can be considered as uniform inside the nanoparticle \cite{Baffou2010}. For the considered metal Ag nanoparticle ensembles embedded in water, the thermal conductivity of metal particle (Ag, $\kappa_{\rm particle} = 429~\rm W\cdot m^{-1} \cdot K^{-1}$) is much higher than that of host medium (water, $\kappa = 0.6~\rm W\cdot m^{-1} \cdot K^{-1}$). The temperature increase $\Delta T_{\rm i}$ (relative to the ambient temperature) inside the ith nanoparticle of the ensemble induced by the incident light yields \cite{Baffou2010}
\begin{equation}
\Delta T_{\rm i} = \sum_{\rm j=1}^{N_{\rm t}}G_{\rm t}(\textbf{r}_{\rm i},\textbf{r}_{\rm j})Q_{\rm j},
\label{T}
\end{equation}
where $Q_{\rm j}=\frac{1}{2}\sigma_{\rm abs}nc\epsilon_0|\textbf{E}^{\rm ext}(\textbf{r}_{\rm j})|^2$, $\sigma_{\rm abs}=k \rm{Im} (\alpha_{\rm j})-\frac{k^4}{6\pi}|\alpha_{\rm j}|^2$ is the absorption cross section of the jth nanoparticle and $n=\sqrt{\epsilon_{\rm m}}$ is the refractive index of the host medium. The thermal Green\textquotesingle s function $G_{\rm t}(\textbf{r}_{\rm i},\textbf{r}_{\rm j})$ with $\textbf{r}_{\rm i}$ different from $\textbf{r}_{\rm j}$ is a scalar Green\textquotesingle s function $G(\textbf{r},\textbf{r}_{\rm j})$  at $\textbf{r}_{\rm i}$, which is associated to the following Poisson equation with a Dirac source distribution $\delta(\textbf{r}-\textbf{r}_{\rm j})$ in an infinite isotropic medium.

\begin{equation}
\nabla \cdot (-\kappa \nabla T(\textbf{r}))=\sum_{\rm j=1}^{N_{\rm t}}Q_{\rm j}\delta(\textbf{r}-\textbf{r}_{\rm j}).
\label{Poisson_T}
\end{equation}

The scalar Green\textquotesingle s function $G(\textbf{r},\textbf{r}_{\rm j})$ is
\begin{equation}
G(\textbf{r},\textbf{r}_{\rm j}) = \frac{1}{4\pi \kappa |\textbf{r}-\textbf{r}_{\rm j}|},
\label{thermal_G}
\end{equation}
where $\kappa$ is the thermal conductivity of the host medium. Hence, for $\textbf{r}_{\rm i}$ different from $\textbf{r}_{\rm j}$,  $G_{\rm t}(\textbf{r}_{\rm i},\textbf{r}_{\rm j})=1/(4\pi \kappa |\textbf{r}_{\rm i}-\textbf{r}_{\rm j}|)$ and for $\textbf{r}_{\rm i}$ equal to $\textbf{r}_{\rm j}$ $G_{\rm t}(\textbf{r}_{\rm i},\textbf{r}_{\rm i})=1/(4\pi \kappa a)$. 

It is worthwhile to mention that the thermal emission by the particles may also affect the temperature increase $\Delta T_{\rm i}$. For a non-absorbing host medium, the net emitted power from the particle j to the thermal bath can be defined as  ${{\mathcal{P}}_{\rm j\leftrightarrow B}^{}}={{\hat{\sigma}}_{\text{abs}}^{}} n^2 {{\sigma }_{B}^{}}(T_{\rm j}^{4}-T_{\text{env}}^{4})$ \cite{Ben2013,Yannopapas2013}, where ${{\hat{\sigma}}_{\text{abs}}^{}}$ is the averaged thermal absorption cross section, $n=\sqrt{\epsilon_{\rm m}}$ is the refractive index, ${\sigma }_{B}^{}$ is Stefan-Boltzmann constant, $T_{\text{env}}$ is the environment temperature, $T_{\text{j}}$ is the temperature of the particle j. For the considered geometry where the power absorbed from the incident light is much more significant than the power exchanged with the environment, ${{\mathcal{P}}_{\rm j\leftrightarrow B}^{}}$ can be neglected with respect to $Q_{\rm j}$ in Eq.~(\ref{T}).

Another important factor that may significantly influence the temperature increase of nanoparticles is the thermal boundary resistance around the particles (e.g., the internal thermal resistance, the interfacial thermal resistance, and the one caused by the molecular coating) \cite{Ali2019JCP,Aksoy2021IJT,Roodbari2022JML,Vera2015IJHMT}, which is well acknowledged to play a key role in nanoparticle-based experiments influencing the heat transfer between the inner plasmonic nanoparticle and the outer surrounding host medium \cite{Alper2010}. For the thermal boundary resistance (TBR) caused by the molecular coating, it has been demonstrated that this kind of TBR indeed can affect the temperature inside the nanoparticle, however, such TBR would not change the temperature distribution in the host medium \cite{Baffou2010}. In addition, in Ref. \cite{Ali2019JCP}, the authors have demonstrated that the internal thermal resistance for the silver nanoparticle embedded in the water is negligible. As for the interfacial thermal resistance, we analyzed its effect on the temperature by applying the spherical heat transfer model and found that this kind of TBR has a similar effect on the temperature as that of the molecular coating, which will be discussed  in the Appendix in detail.

Though different kinds of TBR can exist at the water/silver nanoparticle interface, all of these kinds of TBR will only bring a jump to the temperature inside the nanoparticles and will not change the temperature outside the nanoparticles, which is what usually matters while studying light-induced phenomena. The temperature outside nanoparticles is only dependent on the power $Q$ released by the nanoparticle absorbed from the incident light and the thermal conductivity of the surrounding medium.
That is to say that the temperature in the surrounding medium is dependent only on the multiple scattering and the thermal accumulation and will not change no matter whether there are TBRs or not. To analyze the relation of the multiple scattering and the thermal accumulation on the temperature in the surrounding medium, we take a simple case that neglecting the TBRs, which will not change the main conclusions in this present work concerning the multiple scattering and the thermal accumulation.

\section{Results and discussion}
\label{Results}
The photothermal behavior of two-dimensional nanoparticle ensembles are analyzed with particular focus on the multiple scattering and thermal accumulation effects. Nanoparticle radius ($a$) is 5 nm (diameter of 10 nm). Nanoparticles are composed of metal Ag. The bulk dielectric function of Ag given in Ref.~\cite{Palik} is used in this work, which has been experimentally demonstrated valid for the Ag nanoparticle of the diameter down to at least 7 nm \cite{Hodak1998}. It\textquotesingle s worthwhile to mention that it is easy to include the confinement effect in the present theory only by changing the nanoparticle permittivity. Incident light wavelength is fixed at 383 nm, which is corresponding to the plasmonic resonance of Ag nanoparticle embedded in water with the relative permittivity of $\epsilon_{\rm m} $ = 1.77 and thermal conductivity of $\kappa = 0.6~\rm W\cdot m^{-1} \cdot K^{-1}$. Though we only focus on the multiple scattering effect at resonance frequency in this work, it is noted that effect of multiple scattering on the photothermal behavior is wavelength-dependent. We discussed the multiple scattering effect on photothermal behavior by considering only water as a host medium in this work. As indicated by our previous work \cite{Luo2020ETC}, the relative permittivity of the host medium significantly affects the nanoparticle polarizability, which is strongly relevant to the temperature increase of nanoparticles determined by Eq.~(\ref{T}) combined with  $\sigma_{\rm abs}=k \rm{Im} (\alpha_{\rm j})-\frac{k^4}{6\pi}|\alpha_{\rm j}|^2$ and $Q_{\rm j}=\frac{1}{2}\sigma_{\rm abs}nc\epsilon_0|\textbf{E}^{\rm ext}(\textbf{r}_{\rm j})|^2$. Thus, by tailoring host medium relative permittivity, one may control the photothermal behavior of nanoparticles consequently. The incident light intensity $I_{0}$ is 0.5 mW$\cdot\mu$m$^{-2}$. The lattice spacing $L$ is at least three times larger than the particle radius, which makes the dipole approximation valid \cite{Ben2011,DongPrb2017,Ben2019}. The dimensionless parameters $X$ and $Y$ are defined as $x/[L(N-1)]$ and $y/[L(N-1)]$, respectively.

\subsection{Relation between the multiple scattering and thermal accumulation effects}
\label{Relation_OC_collective}
The light-induced thermal behavior of the nanoparticle ensemble is affected by two main factors, i.e., (a) the multiple scattering effect \cite{Siahpoush2018OC}, and (b) the thermal accumulation effect \cite{Baffou2010,Baffou2020LSA}. We clarify the two factors in brief at first and then discuss the relation between them.

\textit{Multiple scattering effect} The multiple scattering effect concerns the light scattering and absorption of the nanoparticles. As mentioned in Sec.~\ref{DDA}, the external field $\textbf{E}^{\rm ext}(\textbf{r}_{i})$ experienced by the ith nanoparticle at $\textbf{r}_{i}$ can be obtained by Eq.~(\ref{E_ext}), where the second term on the right hand is corresponding to the multiple scattering. When the multiple scattering is negligible, the external field $\textbf{E}^{\rm ext}(\textbf{r}_{i})$ is directly equal to the illuminating incident field $\textbf{E}^{\rm inc}(\textbf{r}_{i})$ by neglecting the second term at the right hand in Eq.~(\ref{E_ext}).

\textit{Thermal accumulation effect} The thermal accumulation effect concerns the thermal diffusion process in the nanoparticle ensemble. As mentioned in Sec.~\ref{TDDA}, the temperature increase $\Delta T$ experienced by the ith nanoparticle calculated by Eq.~(\ref{T}) has two contributions: (a) its own heat generation $\Delta T_{S}$ and (b) the heat generated by the rest other $N_{\rm t}-1$ nanoparticles $\Delta T^{\rm ext}$ under illumination in the ensemble, which yields 
\begin{equation}
\Delta T =\Delta T_{S}+\Delta T^{\rm ext},
\label{T_C}
\end{equation}
where $\Delta T_{S}=G_{\rm t}(\textbf{r}_{\rm i},\textbf{r}_{\rm i})Q_{\rm i}$ is the temperature increase of the ith nanoparticle  due to its own heat generation, $\Delta T^{\rm ext}=\sum_{\rm j\neq i}G_{\rm t}(\textbf{r}_{\rm i},\textbf{r}_{\rm j})Q_{\rm j}$ is the temperature increase of the ith nanoparticle due to the heat generated by the rest other $N_{\rm t}-1$ nanoparticles.

The thermal accumulation effect is corresponding to the second term on the right hand of Eq.~(\ref{T_C}). When the thermal accumulation effect is negligible, the second term on the right hand in Eq.~(\ref{T_C}) is negligible. Each nanoparticle in the nanoparticle ensemble can be treated as an isolated hotspot.

\begin{figure} [htbp]
     \centering
     \subfigure [Multiple scattering effect] {\includegraphics[width=0.48\textwidth]{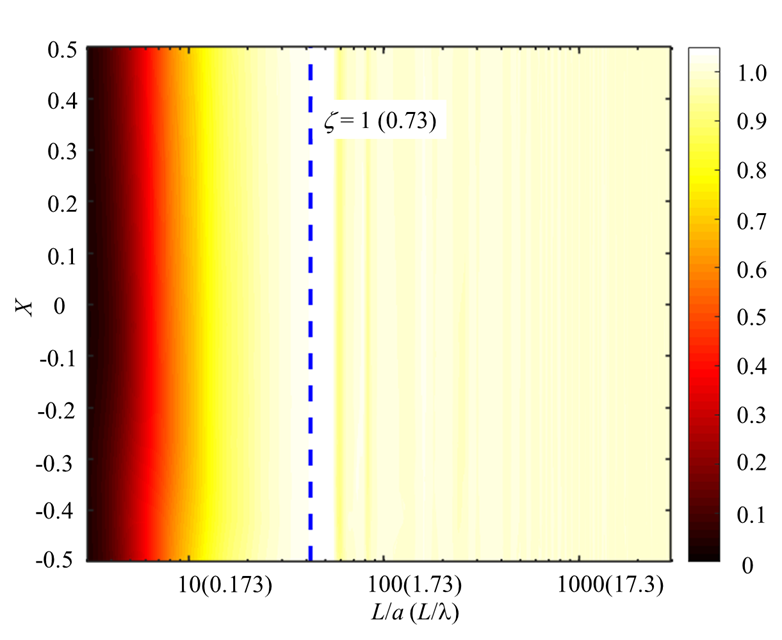}}
     \hspace{8pt}
     \subfigure [Thermal accumulation effect] {\includegraphics[width=0.48\textwidth]{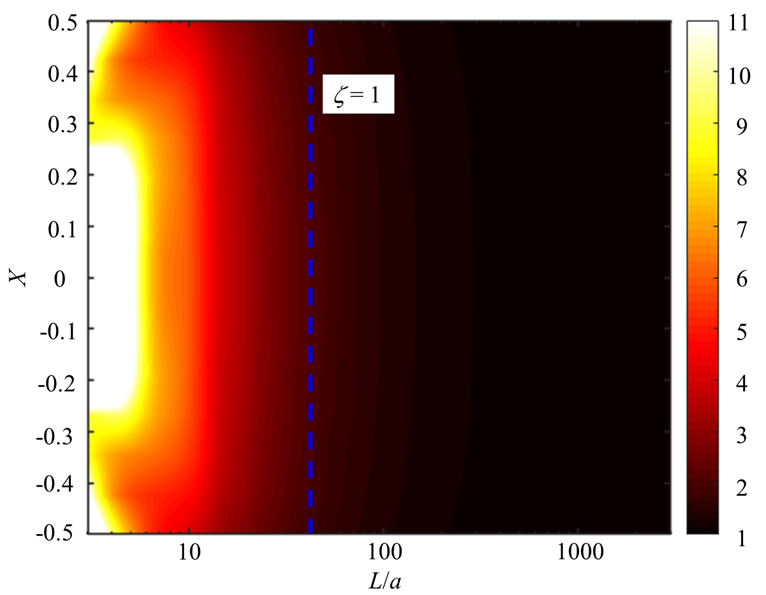}}
        \caption{Dependence of the parameter $\varphi$ on the lattice spacing and dimensionless $X$. (a) We define a parameter $\varphi=\Delta T/\Delta T_{S}$ to evaluate the thermal accumulation effect. $\Delta T_{S}$ is the temperature increase induced by its own heat generation. (b) We define a parameter $\varphi=\Delta T/\Delta T_{0}$ to evaluate the multiple scattering effect. $\Delta T_{0}$ is the temperature increase without the multiple scattering effect. Nanoparticle number is $N=14$ for each lateral edge. The blue-dash line corresponding to the dimensionless parameter defined in Ref. \cite{Baffou2020LSA} $\zeta=L/(3aN)=1$ is added for reference.}
        \label{relation}
\end{figure}

To quantitatively analyze the multiple scattering and the thermal accumulation effect, a parameter $\varphi$ is defined as follows:

\begin{equation}
\varphi=\Delta T/\Delta T_{\rm \nu=0~or~S},
\label{phi}
\end{equation}
where $\Delta T_{\rm \nu=0~or~S}$ are corresponding to the temperature increase without the multiple scattering effect and the temperature increase due to only its own heat generation, respectively. We show the dependence of $\varphi$ for both multiple scattering effect and thermal accumulation effect on two parameters 1) the dimensionless lattice spacing $L/a$ and 2) the dimensionless $X$ in Fig.~\ref{relation}. We considered $14\times 14$ square-lattice nanoparticle ensembles. The blue-dash line corresponding to the dimensionless parameter defined in Ref.\cite{Baffou2020LSA} $\zeta=L/(3aN)=1$ is added for reference. $\zeta >> 1$ is for the negligible thermal accumulation effect. $\zeta << 1$ is for the strong thermal accumulation effect.

As shown in Fig.~\ref{relation}(a), regions with three different $\varphi=\Delta T/\Delta T_{0}$ can be clarified: $\varphi<1$, $\varphi\approx 1$ and $\varphi >1$, respectively. The multiple scattering significantly inhibits $\Delta T$ in the region where $\varphi=\Delta T/\Delta T_{0} <1$, where the lattice spacing is relatively short and the nanoparticle ensemble is relatively dense. The multiple scattering enhances $\Delta T$ in the region $\varphi=\Delta T/\Delta T_{0}>1$. We can see that the enhancement region is right around the blue line. The multiple scattering effect is negligible for the most region, where $\varphi=\Delta T/\Delta T_{0}\approx 1$. We also observe that there are some separated regions where the multiple scattering inhibits $\Delta T$ embedded in the region where the multiple scattering is negligible. In a previous study, the 3D random distribution has been considered \cite{Siahpoush2018OC}. The authors reported that the multiple scattering in the nanoparticle ensembles can significantly inhibit the photothermal behavior without telling the condition when such multiple scattering is negligible from the nanoparticle ensemble structure point of view. As shown in Fig.~\ref{relation}(a), when the new-introduced parameter $\varphi$ approaches to 1, the multiple scattering starts to be less important. When the lattice space $L$ becomes comparable to or even larger than tens of $a$, then multiple scattering inside the nanoparticle ensembles becomes negligible consequently.

As shown in Fig.~\ref{relation}(b), $\varphi=\Delta T/\Delta T_{S}\geq 1$ for the whole region. The region $\varphi=\Delta T/\Delta T_{S}\approx 1$ is the region where the thermal accumulation effect is negligible. The negligible thermal accumulation effect region is consistent with the region $\zeta \gg 1$. For the strong thermal accumulation effect region ($\zeta \ll 1$), we can see  $\varphi \gg  1$, the thermal accumulation effect is strongly in favor of the $\Delta T$. It is noted that the condition $\zeta \gg 1$ for the region of negligible thermal accumulation effect defined in Ref.~\cite{Baffou2020LSA} is not an explicit regime border description. However, the condition that new-introduced dimensionless parameter $\varphi \approx 1$ clearly figures out the regime border of the thermal accumulation explicitly, where the lattice spacing is around 600$a$. When the lattice spacing increases to 600$a$, then each nanoparticle in the ensemble works as a hotspot separately without any influence from nearby nanoparticles and the thermal accumulation effect becomes negligible consequently.

The multiple scattering effect can compete with the thermal accumulation effect for the dense nanoparticle ensembles and can also cooperate with the thermal accumulation effect to enhance $\Delta T$. For the loose enough ensembles, both the multiple scattering effect and the thermal accumulation effect are negligible and the plasmonic nanoparticle can be treated as the isolated hotspot safely.

\subsection{Multiple scattering effect on photothermal behavior}
\label{Many_body_interaction_on_delta_T}
In this section, the optical plasmonic coupling effects on the photothermal behavior of the 2D finite-size square-lattice nanoparticle ensemble are analyzed. It\textquotesingle s worthwhile to mention that the optical plasmonic coupling inhibits the temperature increase of the 3D randomly distributed nanoparticle ensemble \cite{Siahpoush2018OC}.

\begin{figure} [htbp]
\centerline {\includegraphics[scale=0.44]{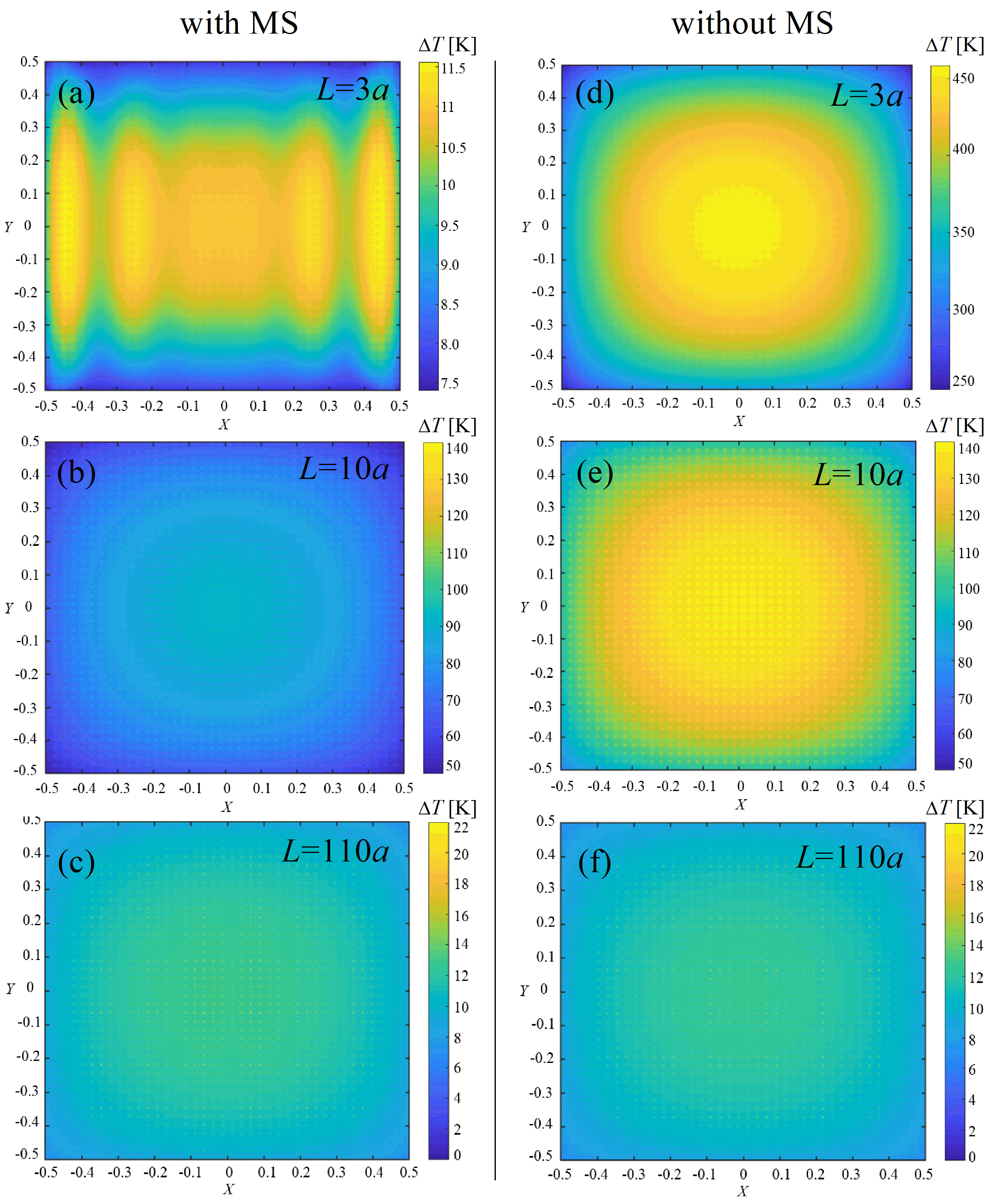}}
        \caption{The temperature increase distribution for 2D ensemble with several different lattice spacings ($L=3a, 10a$ and $110a$, respectively): (a)-(c) $\Delta T$ with multiple scattering effect, (d)-(f) $\Delta T$ without multiple scattering effect. The number of nanoparticles for each lateral edge $N=40$. The dimensionless parameters $X=x/[L(N-1)]$ and $Y=y/[L(N-1)]$.}
        \label{MBI_no_MBI_T}
\end{figure}

The temperature increase distribution for the 2D ensemble with several different lattice spacings is shown in Fig.~\ref{MBI_no_MBI_T} ($L=3a, 10a$ and $110a$, respectively): (a)-(c) the temperature increase distribution with multiple scattering effect, (d)-(f) the temperature increase distribution without multiple scattering effect. Nanoparticle number for each lateral edge $N=40$.

For dense ensembles ($L=3a$ and $10a$), the temperature increase with multiple scattering as shown in Fig.~\ref{MBI_no_MBI_T}(a) and (b) is much less than that without the multiple scattering as shown in Fig.~\ref{MBI_no_MBI_T}(d) and (e), respectively. For the dilute ensemble ($L=110a$), the temperature increase distribution with multiple scattering as shown in Fig.~\ref{MBI_no_MBI_T}(c) is nearly the same as that without the multiple scattering as shown in Fig.~\ref{MBI_no_MBI_T}(f). The strong multiple scattering significantly inhibits the temperature increase of the nanoparticles in the 2D square-lattice ensemble. When the lattice spacing increases, the multiple scattering significantly decreases, which accounts for the negligible inhibition of the $\Delta T$.

\begin{figure} [htbp]
     \centerline{\includegraphics[width=0.48\textwidth]{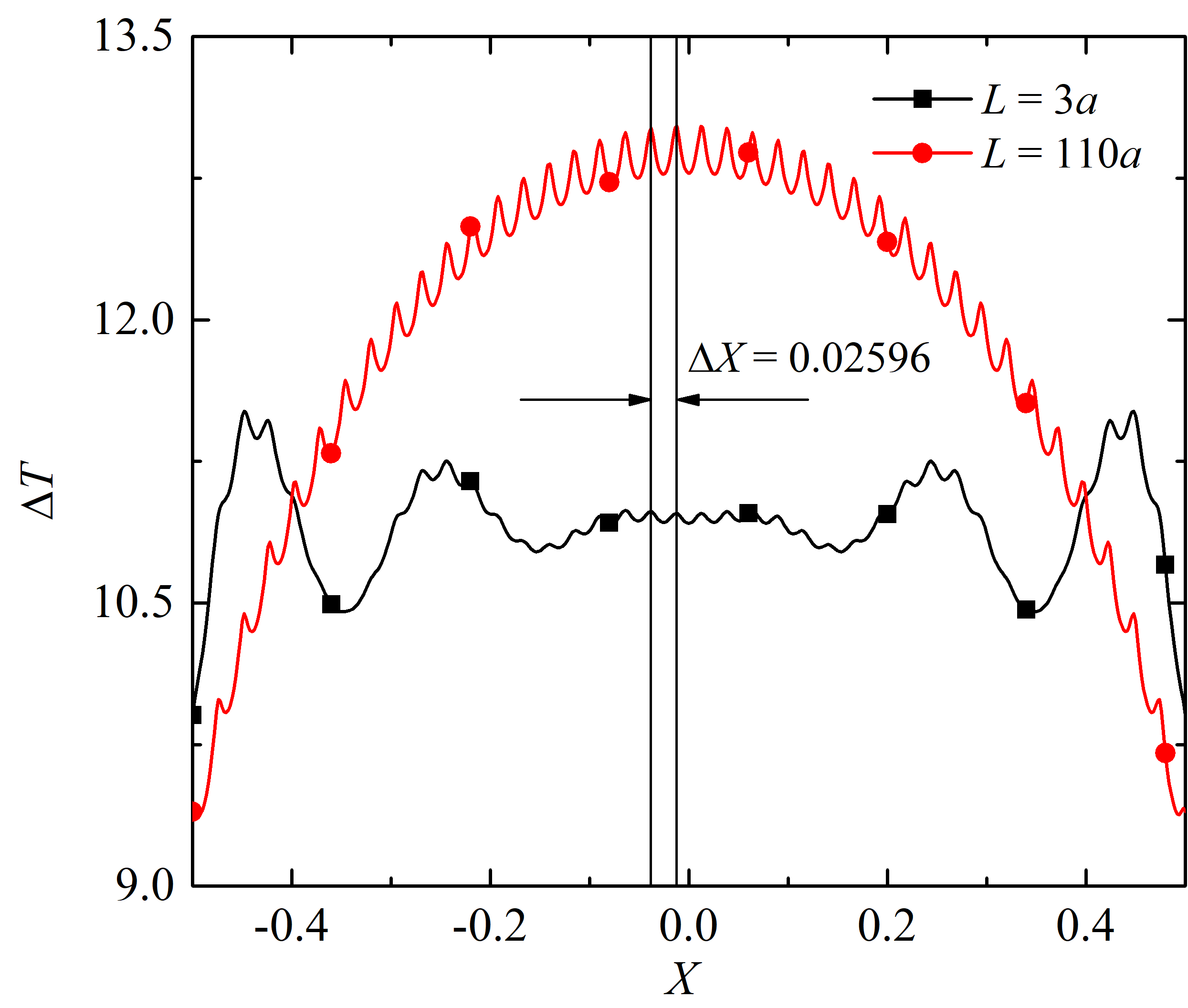}}
        \caption{Temperature increase along the chain of interest with two different lattice spacings ($L=3a$ and $110a$). The separation distance between the neighboring peaks $\Delta X\approx 0.02596$ for the considered two ensembles, which is corresponding to the $\Delta x\approx L$.}
        \label{Temperature_line}
\end{figure}

We extract a chain of nanoparticles of interest near the $x$-axis out of the 2D square-lattice nanoparticle ensemble, as shown in Fig.~\ref{structure}. The chain is defined by $y=L/2$ and $z=0$. The temperature increase along the nanoparticle chains of interest with two different lattice spacings ($L=3a$ and $110a$) is shown in Fig.~\ref{Temperature_line}. The dimensionless and the real separation distance between the neighboring peaks shown in Fig.~\ref{Temperature_line} are $\Delta X\approx 0.02596$ and $\Delta x\approx L$ for the considered two ensembles with two different lattice spacings ($L=3a$ and $110a$), respectively, which is consistent with the reported results \cite{Baffou2013AcsNano}. For the dense ensemble ($L=3a$), we can observe another obvious oscillation of the temperature increase near the boundary along $x$-axis, as shown in Fig.~\ref{MBI_no_MBI_T}(a) and Fig.~\ref{Temperature_line}. For the dilute ensemble ($L=110a$), the temperature increase $\Delta T$ decreases monotonically in general from the center to the boundary along the $x$-axis. The strong multiple scattering in the dense ensemble accounts for the oscillation of temperature increase $\Delta T$ from the center to the boundary of the ensemble along the $x$-axis.

As shown in Fig.~\ref{MBI_no_MBI_T} (a), (b), and (c), it is also noted that the temperature increase for the dense ensemble ($L=3a$) is angle-dependent (seen from the ensemble center), which is quite different from the angle-independent temperature increase for dilute ensembles ($L=10a$ and $110a$). For the dilute nanoparticle ensembles, the multiple scattering becomes less important and thus each particle works as a separate heat source for temperature increase, which results in the angle-independent temperature distribution. The multiple scattering in the dense particle ensemble is expected to be strong and polarization-dependent, which may account for the angle-dependent temperature distribution.  Hence, to some extent, we can tailor the temperature increase distribution by light polarization. However, it\textquotesingle s worthwhile to mention that the light-induced heat transfer behavior in the nanostructures (e.g., dense particle ensembles) is not always polarization-dependent. Recently, a polarization-independent isosbestic temperature increase behavior of nanostructures was proposed \cite{Ma2020AO,Metwally2017}.
. 

\begin{comment}

We show the temperature increase $\Delta T$ distribution for dense ensemble ($L=3a$) illuminated by normal incident light with two different polarizations in Fig.~\ref{polarization_dependence}: (a) $x$ polarization and (b) $y$ polarization. The number of nanoparticles for each lateral edge $N=40$. For the two cases, the only difference is the polarization directions, which are perpendicular to each other. As shown in Fig.~\ref{polarization_dependence}, the temperature increase distributions for the two considered incident lights with mutually perpendicular polarizations are $90^{\circ}$ rotationally symmetrical to each other.

\begin{figure} [htbp]
     \centerline{\includegraphics[width=0.5\textwidth]{polarization_dependence.png}}
        \caption{The temperature increase $\Delta T$ distribution for 2D ensemble ($L=3a$) illuminated by normal incident light with two different polarizations: (a) $x$ polarization and (b) $y$ polarization. The number of nanoparticles for each lateral edge $N=40$. The dimensionless parameters $X=x/L(N-1)$ and $Y=y/L(N-1)$.}
        \label{polarization_dependence}
\end{figure}
\end{comment}

\subsection{Thermal accumulation effect}
\label{Thermal collective effect}
\begin{figure} [htbp]
     \centering
     \subfigure [with multiple scattering] {\includegraphics[width=0.48\textwidth]{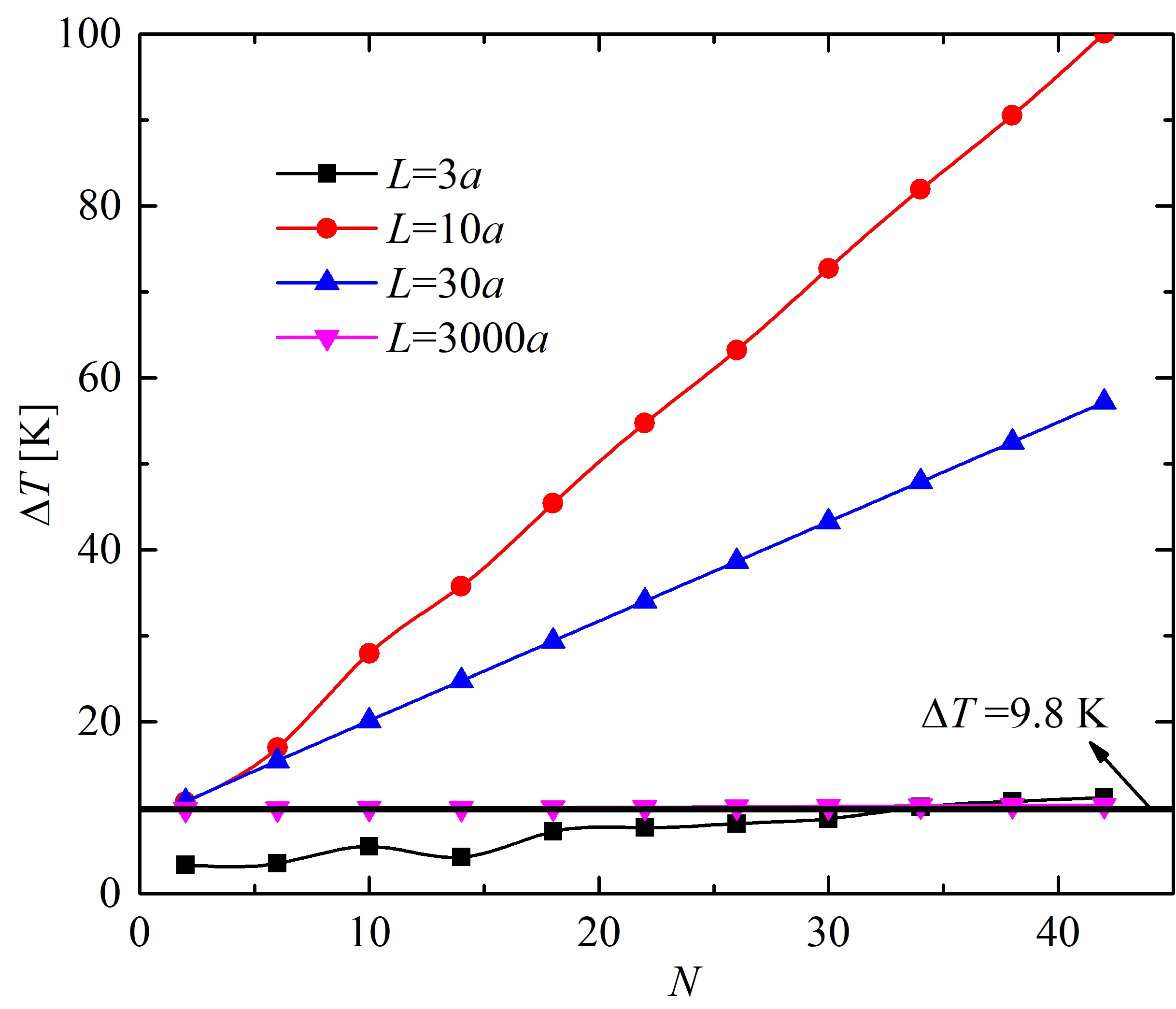}}
     \hspace{8pt}
     \subfigure [without multiple scattering] {\includegraphics[width=0.48\textwidth]{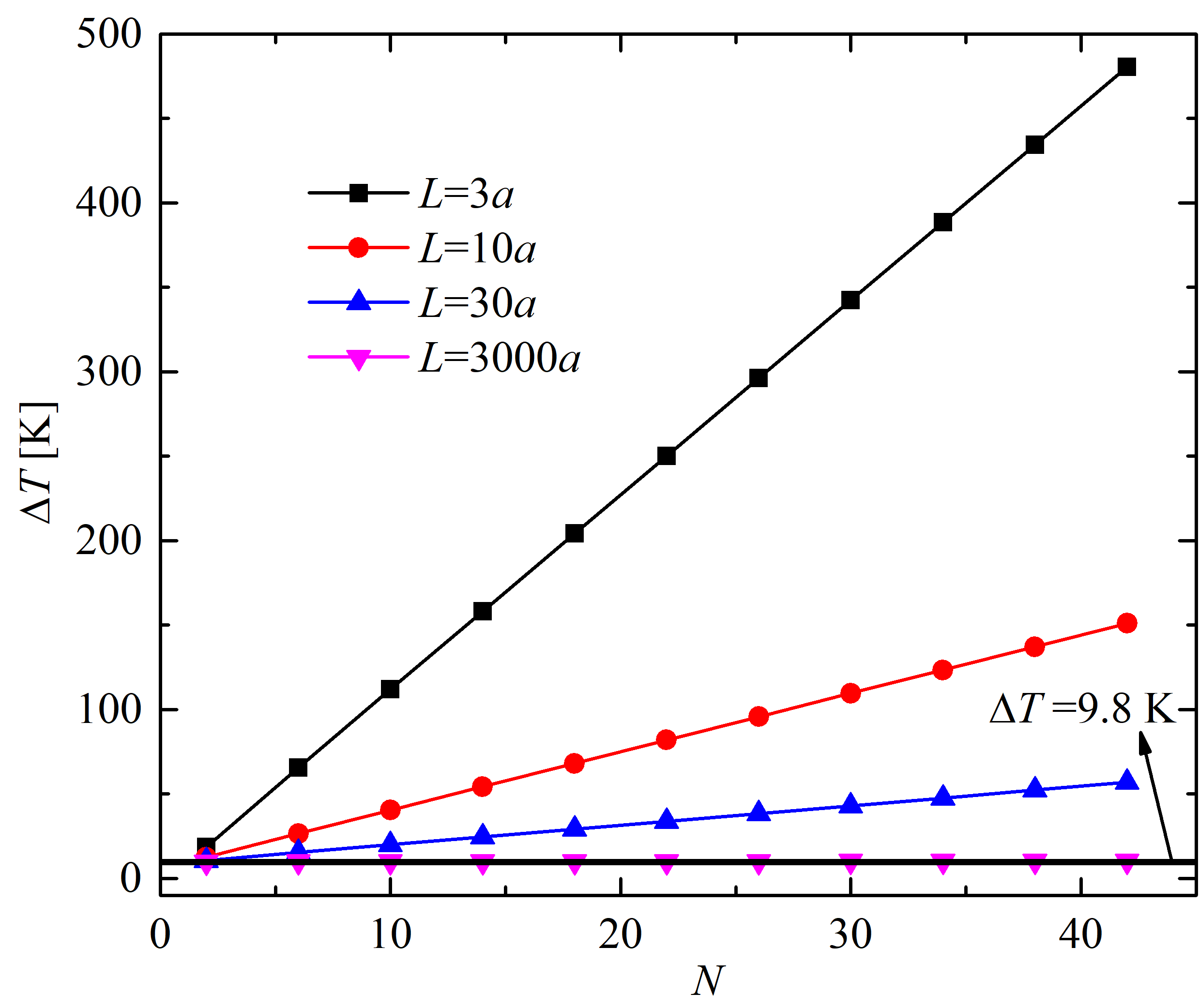}}
        \caption{Dependence of $\Delta T$ of the ensemble center on the nanoparticle number $N$ in each lateral edge: (a) with multiple scattering effect and (b) without multiple scattering effect. Lattice spacing $L=3a,~10a,~ 30a$ and $3000a$. Metal Ag is used. The temperature increase $\Delta T=9.8$ K for the isolated single nanoparticle is also added for reference.}
        \label{collective_number}
\end{figure}

The dependence of $\Delta T$ of the center of the 2D square-lattice ensemble on the nanoparticle number $N$ in each lateral edge is shown in Fig.~\ref{collective_number}: (a) with multiple scattering effect and (b) without multiple scattering effect. Four different lattice spacings are considered, $L=3a,~10a,~ 30a$ and $3000a$, respectively. The temperature increase $\Delta T=9.8$ K for the isolated single nanoparticle is also added for reference.

For the compact ensembles ($L=3a$ and $10a$), $\Delta T$ increases nonlinearly with $N$. While for the loose ensemble ($L=30a$), a linear dependence of $\Delta T$ on $N$ can be observed, which is consistent with the reported results in Ref. \cite{Baffou2010}. For the compact ensemble (e.g., $L=3a$), the temperature increase $\Delta T$ for the central nanoparticle is even less than that of the isolated single nanoparticle. For all the considered ensembles with different lattice spacings, the temperature increase $\Delta T$ without multiple scattering effect increases linearly with $N$, as shown in Fig.~\ref{collective_number}(b). The strong multiple scattering for the compact ensembles accounts for the non-linear dependence of the $\Delta T$ on $N$, observed in Fig.~\ref{collective_number}(a). For the extremely loose ensemble ($L=3000a$), the multiple scattering and thermal accumulation effects are negligible, which accounts for the same temperature increase $\Delta T$ for the central nanoparticle as that of the isolated single nanoparticle.

\begin{figure} [htbp]
     \vspace{8pt}
\centerline {\includegraphics[scale=0.43]{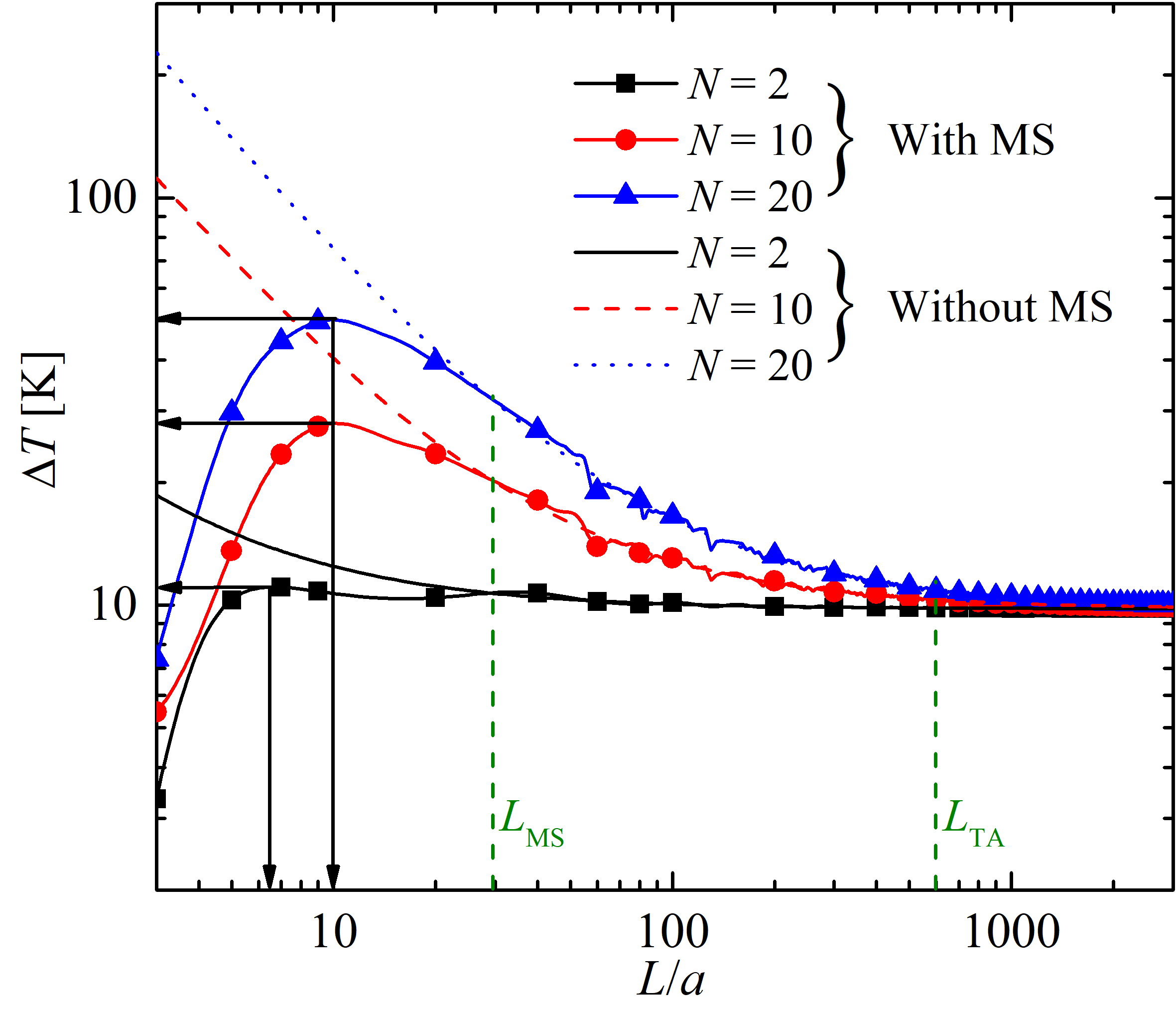}}
        \caption{Dependence of $\Delta T$ for the ensemble center on the lattice spacing $L/a$. $N=2,~10$ and $20$. Both the $\Delta T$ with and without multiple scattering effects are considered. From $L=L_{\rm MS}$, $\Delta T$ with and without multiple scattering effect starts to be identical. From $L=L_{\rm TA}$, $\Delta T$ for ensembles composed of different number of nanoparticles starts to be identical.}
        \label{MBI_lattice_dependence}
\end{figure}

From Fig.~\ref{collective_number}(a) and (b), in general, for an ensemble composed of a certain number of nanoparticles, $\Delta T$ without the multiple scattering effect increases monotonically with the lattice spacing, while $\Delta T$ with the multiple scattering effect increases at first and then decreases with the lattice spacing. The dependence of $\Delta T$ for the ensemble center on the lattice spacing $L/a$ is shown in Fig.~\ref{MBI_lattice_dependence}. Ensembles of three different sizes are considered, $N=2,~10$, and $20$, respectively. Both the $\Delta T$ with and without multiple scattering effects are considered. From $L=L_{\rm MS}$ (the multiple scattering length scale), $\Delta T$ with and without multiple scattering effect starts to be identical, where the multiple scattering is negligible. From $L=L_{\rm TA}$ (the thermal accumulation length scale), $\Delta T$ for ensembles composed of different amounts of nanoparticles starts to be identical, where the thermal accumulation effect is negligible. The length $L_{\rm TA}$ where the thermal accumulation effect is negligible is much larger than the length $L_{\rm MS}$ where multiple scattering starts to be less important. When the thermal accumulation effect is negligible, the multiple scattering also cannot be observed. However, when the multiple scattering effect is negligible, the thermal accumulation effect on photothermal behavior can exist if the lattice spacing $L$ is not large enough.

When considering no multiple scattering effect, the temperature increase $\Delta T$ for the central nanoparticle decreases monotonically with $L/a$. The pure thermal accumulation effect accounts for the monotonic dependence of $\Delta T$ on $L/a$. When considering the multiple scattering effect, $\Delta T$ increases at first and then decreases with $L/a$, as shown with the symbol lines in Fig.~\ref{MBI_lattice_dependence}. The thermal accumulation effect together with the multiple scattering effect account for the non-monotonic dependence of $\Delta T$ on $L/a$.

We also can notice that small oscillations of $\Delta T$ occur at around few tens of $L/a$ in Fig.~\ref{MBI_lattice_dependence}. Note that $a = 5$ nm, so the lattice spacings $L$ corresponding to small oscillations are comparable to or even larger than the considered wavelength (383 nm). According to Zou and Schatz (2004) \cite{Zou2004}, for a fixed wavelength, the extinction efficiency of two-dimensional hexagonal arrays is not monotonously dependent on the lattice spacing, as shown in FIG. 3 of Ref. \cite{Zou2004}. Such non-monotonously-dependent optical properties may account for the small oscillations of $\Delta T$ when increasing $L/a$ for two-dimensional nanoparticle arrays.

\section{Conclusion}
Light-induced heat transfer of the two-dimensional nanoparticle ensembles embedded in water is investigated by means of the Green’s function approach with a focus on proposing a clear regime map of both the multiple scattering and the thermal accumulation. The dimensionless parameter $\varphi$ is defined to quantify the multiple scattering and collective effects on photothermal behavior. For 2D ordered nanoparticle ensemble, similar to the 3D random nanoparticle ensemble, multiple scattering can also inhibit temperature increase $\Delta T$ of nanoparticles. The more compact the nanoparticle ensemble is, the stronger the multiple scattering on thermal behavior is. When the lattice spacing increases to tens of nanoparticle radius, the multiple scattering becomes insignificant. In addition, when $\varphi \approx 1$ and lattice spacing increases to hundreds of nanoparticle radius, the thermal accumulation effects are weak and can be neglected safely. The distribution of temperature increase of nanoparticle is polarization-dependent, especially for the compact nanoparticle ensemble. While for dilute ensemble, such polarization-dependent temperature increase distribution can not be observed. The temperature increase $\Delta T$ of the center of the ensemble increases linearly with increasing the ensemble size without considering multiple scattering. However, when considering multiple scattering, for the compact ensemble, $\Delta T$ increases non-linearly with increasing the ensemble size. For a fixed ensemble size, $\Delta T$ increases non-linearly with increasing the lattice spacing. This work may help for the understanding of the light-induced thermal transport in the 2D particle ensemble.

\begin{acknowledgments}
This work is supported by the National Natural Science Foundation of China (No. 51976045). M.G. Luo also thanks for support from the China Postdoctoral Science Foundation (2021M700991). In addition, we acknowledge greatly the kind and helpful suggestions from the two anonymous reviewers.
\end{acknowledgments}

\appendix*
\section{Influence of the interfacial thermal resistance on the temperature profile} 
\label{TBRs_effects}
We applied a spherical heat transfer model to investigate the effect of the interfacial thermal resistance between nanoparticle and water on the temperature increase. Considering that the thermal conductivity of the nanoparticle $\kappa_{\rm particle} \gg \kappa $ (the conductivity of the surrounding medium), the temperature inside nanoparticle ($r < a$) is uniform $T_0$. The general fomula of the temperature outside the nanoparticle ($r > a$) is $T(r)=C_1/r+C_2$, where the constants $C_1$ and $C_2$ can be obtained by applying the boundary conditions. The interfacial thermal resistance is $R$.  The power absorbed by the nanoparticle with radius $a$ from incident light is $Q$, which is released to the surround medium. We have the following boundary conditions.
\begin{equation}
\begin{aligned}
& T(r\rightarrow \infty)=0,
\\& \kappa (\frac{C_1}{r^2})=\frac{Q}{4\pi r^2}.
\label{BC}
\end{aligned}
\end{equation}
We can obtain that $C_1=\frac{Q}{4\pi \kappa}$ and $C_2=0$. Hence, the temperature outside the nanoparticle yields
\begin{equation}
T(r)=\frac{Q}{4\pi \kappa r}.
\label{T_out}
\end{equation}

At the interface of nanoparticle/water, we have the following relation:
\begin{equation}
\frac{T_0-T(r)|_{r\rightarrow a}}{R}=\frac{Q}{4\pi a^2}.
\label{T_inside1}
\end{equation}
Therefore, the temperature inside the nanoparticle is
\begin{equation}
T_0=\frac{Q}{4\pi \kappa a}+\frac{QR}{4\pi a^2}.
\label{T_inside}
\end{equation}

The whole temperature profile yields
\begin{equation}
\begin{aligned}
& r<a; T_0=\frac{Q}{4\pi \kappa a}+\frac{QR}{4\pi a^2},
\\& r\geq a; T(r)=\frac{Q}{4\pi \kappa r}.
\label{T_profile}
\end{aligned}
\end{equation}

From the above Eq.~(\ref{T_profile}) for the temperature profile of the whole domain, the interfacial thermal resistance $R$ will not change the temperature outside the nanoparticles ($r \geq a$), which usually matters when studying the light-induced phenomena, similar to that observed for the coating-induced thermal resistance reported in Ref.~\cite{Baffou2010}. The interfacial thermal resistance $R$ may significantly affect the temperature inside the nanoparticles, which will bring an additional temperature increase $\frac{QR}{4\pi a^2}$ as compared to the previous one by neglecting the interfacial thermal resistance. Especially, for a small particle with a large interfacial thermal resistance $R$, the temperature jump across the nanoparticle/water interface will be significant.

%\bibliography{Photothermal}

\providecommand{\noopsort}[1]{}\providecommand{\singleletter}[1]{#1}%

\end{document}